\documentclass[prl,twocolumn,superscriptaddress]{revtex4}

\usepackage{hyperref}
\usepackage{rotating}
\usepackage{graphicx}
\usepackage{latexsym}
\usepackage{amssymb}
\usepackage{amsfonts}
\usepackage{soul}
\usepackage{color}
\usepackage{amsmath}
\usepackage{color}
\usepackage{lipsum}
\usepackage{natbib}

\usepackage{changes}
    \colorlet{Changes@Color}{blue}

\newcommand{\be}{\begin{equation}}
\newcommand{\ee}{\end{equation}}
\newcommand{\bes}{\begin{equation*}}
\newcommand{\ees}{\end{equation*}}
\newcommand{\bea}{\begin{eqnarray}}
\newcommand{\eea}{\end{eqnarray}}
\newcommand{\ba}{\begin{array}}
\newcommand{\ea}{\end{array}}

\newcommand{\Tr}{\mathrm{Tr}}

\newcommand{\dagga}{{\phantom{\dagger}}}
\newcommand{\bw}{\begin{widetext}}
\newcommand{\ew}{\end{widetext}}
\newcommand{\changes}{\textcolor{black}}
\newcommand{\change}{\textcolor{black}}

\begin{document}

\title{Entanglement \change{properties} and ground-state statistics of free bosons}

\author{Luca Dell'Anna}
\affiliation{Department of Physics and Astronomy, University of Padova, Via F. Marzolo 8, 35131 Padova, Italy}
\affiliation{Padua Quantum Technologies Research Center, University of Padova, Via Gradenigo 6, Padova, Italy}
\begin{abstract}
We calculate analytically the entanglement and R\'enyi entropies, the negativity and the mutual information together with all the density and many-particle correlation functions for free bosons on a lattice in the ground state, for both homogeneous and inhomogeneous systems. We show that all those quantities can be derived from a multinomial form of the reduced density matrix in the configuration space whose diagonal elements dictate the statistics of the particle distribution, while the off-diagonal coherence terms control the quantum fluctuations. 
We provide by this analysis a unified approach based on a reduced density matrix technique useful to calculate both the entanglement properties and an infinite number of correlation functions. 
\end{abstract}

\maketitle
\section{Introduction}
In spite of its simplicity the system of non-interacting bosons placed on a $d$-dimensional lattice represents a very peculiar case having particle number fluctuations, within a subsystem $A$, which obey a volume law while the entanglement entropy scales logarithmically \cite{peschel1,peschel,klich,ding,casini,laflorencie}, violating the so-called area law \cite{eisert}, 
\changes{in the same way as in the ferromagnetic Heisenberg model \cite{popkov, salerno, alvaredo, alba1} and analogously to conformal field theories \cite{vidal1,calabrese}, random \cite{rafael} or unconventional quantum spin chains \cite{luca_sc}. }  

\change{Most of the studies devoted to the entanglement properties of bosonic systems are focused on relativistic free bosonic field theories, which can be mapped to coupled harmonic oscillators \cite{peschel1,peschel,casini,laflorencie,eisert} with an infinite number of modes. We will consider, instead, the entanglement properties and the ground-state statistics of systems made of a fixed number of free non-relativistic bosons which form a condensate on a lattice. In this case,} \changes{the entanglement entropy and the particle number fluctuations} 
can be explained by assuming a binomial distribution of the particles inside the subsystem under study \cite{klich,ding,alba,roscilde}. 
However this distribution is verified also for a confined non-extended system like bosons in a double well \cite{luca1}.

In order to unveil the correct particle distribution in an extended region one has to consider moments of higher order than the variance of the number of particles finding, in this way, the full statistics in the ground state. 
\changes{This state is a Bose-Einstein condensate which can be described by a single-particle wavefunction. In the homogeneous case, all sites are equivalent, therefore geometry and dimension do not play any role in the bipartition of the system, only the number of sites in the subsystems really matter.}

We show that the reduced density matrix is once again the basic tool containing all the informations needed to explain either the entanglement properties of our quantum system and the many-particle correlation functions. 

We calculate the complete reduced density matrix in the configuration space after a bipartition of the full system into two subsets and derive analytically the entanglement entropy, the R\`enyi entropies, the quantum negativity \cite{lewe,vidal} and the mutual information \cite{groi,wolf} between two separated regions of a subsystem. 
We show that, in the thermodynamic limit, the negativity vanishes  whereas the mutual information is always finite, independently from the distance between the two regions, suggesting that the mixed state describing the subsystem is an entangled positive partial transpose (PPT) state. 

At the same time we show that the reduced density matrix can be seen as a statistical distribution useful to calculate analytically all the correlation functions written in terms of statistical averages. 
In the thermodynamic limit, looking only at the density correlation functions, the system of free bosons behaves like an uncorrelated gas 
since density operators are diagonal in the configuration space. 
All the density correlations, therefore, can be calculated exactly at any order from the moment generating function of a multinomial distribution. 
The coherence among the particles, instead, is the origin of off-diagonal terms in the reduced density matrix and is disclosed by the two-point single-, pair and many-particle correlation functions, which are also calculated exactly at any order.

 \section{Model and ground state}
The model describing free bosons hopping on a generic $d$-dimensional lattice is the following
\begin{equation}
\label{H1}
H=-\sum_{ij}\, t_{ij} b_i^\dagger b^\dagga_j 
\end{equation}
where $t_{ij}$ is the positive-definite hopping amplitude for a particle to jump from a site $i$ to a site $j$ of the lattice with $V$ sites while $b_i, b_i^\dagger$ are the bosonic annihilation and creation operators. We will consider in particular a translationally invariant system with periodic boundary condition in $d$ dimensions, namely free bosons on a $d$-torus. 
At zero temperature, all particles are in the ground state, which, for a system made by $N$ particles in $V$ sites ($V$ is the volume), is given by
\bea
\nonumber &&
|\Psi\rangle=\frac{1}{\sqrt{N!}}\Big(\sum_j \xi_j b_j^\dagger\Big)^N  \hspace{-0.cm}|0\rangle\\
&&\phantom{|\Psi\rangle}
=\sum^N_{\{n_1\dots n_V\}}\frac{\sqrt{N!}}{\prod_{i=1}^V n_i!}\prod_{j=1}^{V}\xi_j^{n_j}b_j^{\dagger{n_j}}|0\rangle
\label{state}
\eea
where $\xi_j$ are complex numbers representing the single-particle wave-function fulfilling the normalization 
condition $\sum_{j=1}^{V}|\xi_j|^2=1$. 
For an homogeneous system, with periodic boundary conditions, one has $\xi_j=1/\sqrt{V}$. 
The sum over $\{n_1,\dots,n_V\}$, means the sum over all the configurations $(n_1,\dots,n_{V})$ of the occupation numbers 
with the constraint $\sum_{i=1}^V n_i=N$, namely the summation notation means 
\[\sum^N_{\{n_1\dots n_V\}}\equiv
\sum_{n_1=0}^N\sum_{n_2=0}^{N-n_1}\dots \sum_{n_{V-1}=0}^{N-\sum_{i=1}^{V-2} n_i}\delta_{N-\sum_{i=1}^{V} n_i}.\]

\section{Reduced density matrix}
Let us now calculate the reduced density matrix 
$\hat\rho$ after partitioning the full system in two blocks, 
$A\equiv [1,{V_A}]$ and $B\equiv [{V_A}+1,V]$, and tracing out the degrees of freedom belonging to the second block,
\bea
\nonumber\hat\rho=\Tr_B(|\Psi\rangle\langle\Psi|)=\hspace{-0.2cm}
\sum^{N\,\prime}_{\{n_{V_A\hspace{-0.03cm}+1},..,n_{V}\}}\hspace{-0.4cm}
\frac{\langle 0_B| \prod_{i={V_A}+1}^{V}b_i^{n_i}|\Psi\rangle}{\sqrt{\prod_{i=V_A +1}^{V}n_i!}} \\
\times \frac{\langle \Psi| \prod_{i={V_A}+1}^{V}b_i^{\dagger n_i}|0_B\rangle}{\sqrt{\prod_{i=V_A +1}^{V}n_i!}}
\label{rdmA}
\eea
where the sum over all the occupation number configurations $(n_{V_A+1},\dots,n_{V})$ fulfilling $\sum_{i=V_A+1}^V n_i\le N$ is denoted by $\sum^{N\,\prime}_{\{n_{V_A+1}\dots n_V\}}\equiv \sum_{N_B=0}^N\sum^{N_B}_{\{n_{V_A\hspace{-0.05cm}+1}\dots n_{V}\}}$ using the previous definition, or can be expressed as 
\[\sum^{N\,\prime}_{\{n_{V_A+1}\dots n_V\}}
\equiv \sum_{n_{V_A+1}=0}^N \,\sum_{n_{V_A+2}=0}^{N-n_{V_A+1}}\dots\sum_{n_V=0}^{N-\sum_{i=V_A+1}^{V-1} n_i}\] and where the vacuum state is split as $|0\rangle=|0_A\rangle|0_B\rangle$. 
Performing the trace over $B$ we get
\bea
\nonumber \hat\rho=\hspace{-0.05cm}\sum_{N_A=0}^N
\sum^{N_A}_{\{n_1\dots n_{V_A}\hspace{-0.05cm}\}}\hspace{-0.05cm}
\sum^{N_A}_{\{m_1\dots m_{V_A}\hspace{-0.05cm}\}}
\hspace{-0.2cm}\frac{N!\,\Big(1-\sum_{i=1}^{V_A} |\xi_i|^2\Big)^{N-N_A}}{\left(N-N_A\right)!\,\prod_{i=1}^{V_A}n_i!\, m_i!}\\
\times\,\prod_{j=1}^{V_A}\xi_j^{n_j}b_j^{\dagger n_j}|0_A\rangle \langle 0_A|\prod_{j=1}^{V_A}\xi_j^{*m_j}b_j^{m_j}\hspace{0.3cm}
\label{rdmA2}
\eea
This reduced density matrix is a $[(N+{V_A})!/({V_A}!N!)]\times [(N+{V_A})!/({V_A}!N!)]$ block diagonal matrix which can be written as 
\be
\label{rho_matrix}
\hat\rho=\left(\begin{array}{ccccc}
\hat\rho_N&0&\dots&0\\
0&\hat\rho_{N-1}&\dots&0\\
\vdots&\vdots&\ddots&\vdots\\
0&0&\dots&\hat\rho_{0}\\
\end{array}
\right)
\ee
where each block $\hat\rho_{N_A}$, with $0\le N_A\le N$,  is a $[(N_A+V_A-1)!/(N_A!(V_A-1)!)]\times [(N_A+V_A-1)!/(N_A!(V_A-1)!)]$ fully sparse matrix 
whose elements, in the configuration space, are
\bea
\label{rdm}
\rho_{\substack{n_1\dots n_{V_A}\\m_1\dots m_{V_A}}}=
\frac{N!\,\Big(1-\sum_{i=1}^{V_A} |\xi_i|^2\Big)^{N-N_A}
}{\left(N-N_A\right)!}\prod_{i=1}^{V_A}
\frac{\xi_i^{n_i}\xi_i^{* m_i}}{\sqrt{n_i! \,m_i!}}
\eea
with  $N_A=\sum_{i=1}^{V_A} n_i=\sum_{i=1}^{V_A} m_i$. The null elements in Eq.~(\ref{rho_matrix}) are those 
with $\sum_{i=1}^{V_A} n_i\neq \sum_{i=1}^{V_A} m_i$. \\
It is convenient to write explicitly the diagonal elements of Eq.~(\ref{rho_matrix}), denoted for simplicity $\rho_{n_1\dots n_{V_A}}\equiv \rho_{\substack{n_1\dots n_{V_A}\\n_1\dots n_{V_A}}}$, 
\be
\label{rdmD}
\rho_{n_1\dots n_{V_A}}=
\frac{N!\,\Big(1-\sum_{i=1}^{V_A} |\xi_i|^2\Big)^{N-N_A}
}{\left(N-N_A\right)!}\prod_{i=1}^{V_A}
\frac{|\xi_i|^{2n_i}}{n_i!}
\ee
which has the form of a multinomial distribution \cite{luca2}, and will be useful, as we will be seeing in the next sections, in deriving the density  correlation functions. \\
We notice that Eq.~(\ref{rdm}) can be factorized so that, for any $0\le N_A\le N$, we can rewrite it as the outer product of two vectors 
\be
\label{WW}
\hat \rho_{N_A}={\cal C}_{N_A} \vec W\otimes \vec {W}^*={\cal C}_{N_A} \vec W\,\vec {W}^\dagger
\ee
where 
\be
\label{CNA}
{\cal C}_{N_A}=\frac{N!}{(N-N_A)!}\,\Big(1-\sum_{i=1}^{V_A} |\xi_i|^2\Big)^{N-N_A}
\ee
$\vec W$ a vector with dimension $(N_A+V_A-1)!/(N_A!(V_A-1)!)$ in configuration space whose elements are
\be
W_{n_1\dots n_{V_A}}=\prod_{i=1}^{V_A} \frac{\xi_i^{n_i}}{\sqrt{n_i!}}
\ee
The eigenvalues of Eq.~(\ref{WW}) can now be easily calculated. They are all zeros except one which we call $\rho_{N_A}$ (without the hat symbol)  given by
\bea
\rho_{N_A}=\Tr(\hat \rho_{N_A})={\cal C}_{N_A} |\vec W|^2
={\cal C}_{N_A} \hspace{-0.35cm}\sum_{\{n_1\dots n_{V_A}\}}^{N_A}\prod_{i=1}^{V_A} \frac{|\xi_i|^{2n_i}}{n_i!}\hspace{0.2cm}
\eea
Performing the summation and using Eq.~(\ref{CNA}) we get the explicit eigenvalues of the reduced density matrix which have the form of a binomial distribution
\be
\label{rhoNA}
\rho_{N_A}
=\frac{N!\,\Big(1-\sum_{i=1}^{V_A} |\xi_i|^2\Big)^{N-N_A}\Big(\sum_{i=1}^{V_A} |\xi_i|^2\Big)^{N_A}
}{\left(N-N_A\right)!\,N_A!}
\ee
This result is the same as that we could have obtained resorting to the Schmidt decomposition procedure \cite{ding}. The advantage of this approach is that we have the full density matrix in the configuration representation whose diagonal elements, fulfilling
\be
\label{rhosum}
\sum^{N_A}_{\{n_1\dots n_{V_A}\}}\rho_{n_1\dots n_{V_A}}=\rho_{N_A}
\ee
will appear in the calculation of the density correlation functions while, as we will see, the off-diagonal elements, instead, 
are responsible for the many-particle quantum hopping processes. 
\section{Entanglement measures}
After finding and diagonalizing the reduced density matrix we can calculate some entanglement properties. 
\subsection{Entanglement entropy}
The Von Neumann entropy is given by
\be
\label{Sdef}
S=-\Tr_A\left(\hat\rho\ln\hat\rho\right)=-\sum_{N_A=0}^N\rho_{N_A}\ln \rho_{N_A}
\ee
where $\rho_{N_A}$ are the non-null eigenvalues of the reduced density matrix, Eq.(\ref{rho_matrix}), given in Eq.~(\ref{rhoNA}).
After performing the trace we get the general result for the entanglement entropy of a Bose-Einstein condensate, also in the presence of space modulations dictated by the single-particle wavefunction $\xi_i$ inside the subsystem $A$. 
For the homogeneous case we have that the single particle wavefunction is uniform everywhere and given by
\be
\label{filling}
|\xi_i|^2=\frac{1}{V}=\frac{\nu}{N}
\ee
where we introduced the filling fraction $\nu=N/V$. 
In the thermodynamic limit, $N\rightarrow \infty, V\rightarrow \infty$, keeping  $\nu$ finite, since
\bea
&&\lim_{N\rightarrow\infty} \frac{N!}{(N-N_A)!\;N^{N_A}}=1\\
&&\lim_{N\rightarrow\infty} \left(1-\frac{V_A\nu}{N}\right)^{N}=e^{-V_A\nu}
\eea
Eq.~(\ref{rhoNA}) becomes simply
\be
\label{poisson}
\rho_{N_A}=\frac{1}{N_A!}\,(V_A\nu)^{N_A} e^{-{V_A} \nu}
\ee
In the same limit the diagonal elements of $\hat \rho$, Eq.~(\ref{rdmD}), become
\be
\label{poissonD}
\rho_{n_1\dots n_{V_A}}=\frac{1}{\prod_{i=1}^{V_A} n_i!}\,\nu^{\left(\sum_{i=1}^{V_A} n_i\right)} e^{-{V_A} \nu}
\ee
which clearly fulfill Eq.~(\ref{rhosum}). 
From Eqs.~(\ref{Sdef}) and (\ref{poisson}) we get the following result for the entanglement entropy
\be
\label{entropy}
S={V_A}\nu \big(1-\ln(V_A\nu)\big)+e^{-V_A\nu}\sum_{n=0}^{\infty}\frac{(V_A\nu)^{n}}{n!}\ln n!
\ee
For $V_A\ll \nu^{-1}$ the entropy is $S\simeq {V_A}\nu \big(1-\ln(V_A\nu)\big)$. 
For very large values of the filling fraction $\nu$ ($N\gg V$), or for large subsystems, namely for $V_A\nu\gg 1$, Eq.~(\ref{entropy}) becomes 
\be
\label{Sgauss}
S\simeq \frac{1}{2} \ln(2\pi e V_A\nu)
\ee
which can be obtained approximating the last term in Eq.~(\ref{entropy}) as follows
\[
e^{-V_A\nu}\sum_{n=0}^{\infty}\frac{(V_A\nu)^{n}}{n!}\ln n!=\langle\ln N_A!\rangle_{\rho_{N_A}}
\hspace{-0.1cm}\sim \ln ([V_A\nu]!)+\frac{1}{2}
\]
where $\langle\ln N_A!\rangle_{\rho_{N_A}}$ is a statistical average with weight Eq.~(\ref{poisson}), 
and using the Stirling series 
\[
\ln ([V_A\nu]!)\sim V_A\nu\,(\ln (V_A\nu)-1)+\frac{1}{2}\ln(2\pi V_A\nu)
\]
Actually, in the latter situation, $V_A\nu\gg 1$, the Poisson distribution Eq.~(\ref{rhoNA}) turns to a Gaussian one 
\be
\label{rhogauss}
\rho_{N_A}\simeq \frac{e^{-\frac{(N_A-\nu V_A)^2}{2 \nu V_A(1-\nu V_A/N)}}}{\sqrt{2\pi \nu V_A(1-\nu V_A/N)}} 
\ee
so that one can easily verify that in the thermodynamic limit the entropy is given by Eq.~(\ref{Sgauss}). In the generic case, for an inhomogeneous system Eq.~(\ref{Sgauss}) can be generalized as follows
\be
\label{Sgauss2}
S\simeq \frac{1}{2} \ln\Big[2\pi e N
\Big(\sum_{i=1}^{V_A} |\xi_i|^2\Big)\Big(1-\sum_{i=1}^{V_A} |\xi_i|^2\Big)
 \Big]
\ee
obtained putting $N\sum_{i=1}^{V_A} |\xi_i|^2$ instead of $\nu V_A$ in Eq.~(\ref{rhogauss}).
\changes{Before we proceed a comment is in order. As already discussed in Ref.~\cite{ding}, the standard approach described in Refs.~\cite{peschel1,peschel} for calculating the entanglement entropy through the two-point correlation function matrix 
$C_{ij}=\langle b_i^\dagger b_j\rangle$, truncated within the subsystem $A$, does not lead to the correct result for a Bose-Einstein condensate. By that procedure the entanglement entropy reads 
$S=\sum_{\ell}\left[(1+c_\ell)\ln(1+c_\ell)-c_\ell\ln c_\ell\right]$ with $c_\ell$ the eigenvalues of $C_{ij}$. For the homogeneous case 
there is only one non-zero eigenvalue of $C_{ij}$, namely $c_\ell=\nu V_A\delta_{\ell 0}$, therefore, for large $V_A$ the entropy is $S\simeq \ln(V_A\nu)$, which differs from Eq.~(\ref{Sgauss}) in the prefactor. This discrepancy originates from the fact that the approach used in Refs. \cite{peschel1,peschel} relies on the conventional formulation of the Wick's theorem for correlation functions, which cannot be applied for the ground-state in Eq.~(\ref{state}), as we will be seeing in what follows (see Eqs.~(\ref{c1}) and (\ref{4p})).}

\subsection{R\'enyi entropies}
Let us calculate the R\'enyi entopies of generic order $\alpha$ 
 defined by
\be
S_\alpha=\frac{1}{1-\alpha}\ln \Tr({\hat\rho}^\alpha)
\ee
which, after diagonalizing the reduced density matrix $\hat \rho$, becomes
\be
S_\alpha=\frac{1}{1-\alpha}\ln \Big[\sum_{N_A=0}^{N}\rho_{N_A}^\alpha\Big]
\ee
where $\rho_{N_A}$ is given in Eq.~(\ref{rhoNA}). In the Gaussian regime, namely for $N\sum_{i=1}^{V_A} |\xi_i|^2\gg 1$, we can calculate $S_\alpha$ analytically 
using the following result
\bea
\nonumber 
&&\Tr({\hat\rho}^\alpha)\simeq \int_{-\infty}^\infty dx \left(\frac{e^{-\frac{(x-NP)}{2 NP(1-P)}}}{\sqrt{2\pi NP(1-P)}}\right)^\alpha\\
&& \phantom{\Tr({\hat\rho}^\alpha)}=\left(\frac{2\pi NP(1-P)}{\alpha}\right)^{\frac{1-\alpha}{2}}
\eea
where $P=\sum_{i=1}^{V_A} |\xi_i|^2$, getting
\be
S_\alpha\simeq \frac{1}{2} \ln\Big[2\pi \alpha^{\frac{1}{\alpha-1}} N
\Big(\sum_{i=1}^{V_A} |\xi_i|^2\Big)\Big(1-\sum_{i=1}^{V_A} |\xi_i|^2\Big)
 \Big]
\ee
Notice that in the limit $\alpha\rightarrow 1$ we recover the von Neumann entropy reported in Eq.~(\ref{Sgauss2}). 
In the homogeneous case and in the thermodynamic limit $NP(1-P)\rightarrow V_A\nu$ and we get
\be
\label{renyi}
S_\alpha\simeq \frac{1}{2} \Big[ \ln\Big(2\pi V_A\nu\Big)+ \frac{\ln \alpha}{\alpha-1} \Big]
\ee
which is the generalization of Eq.~(\ref{Sgauss}). We notice that the minimum R\'enyi entropy obtained for $\alpha\rightarrow \infty$ in Eq.~(\ref{renyi}), known as min-entropy, is finite, $\frac{1}{2} \ln\big(2\pi V_A\nu\big)$.

\subsection{Negativity}
Let us assume now to split the system $A$ in two parts $A_1$ and $A_2$ with number of sites $V_1$ and $(V_A-V_1)$, respectively. We will calculate the quantum negativity measured after partial transposition of the second block, $A_2$. It is convenient to rewrite Eq.~(\ref{rdmA2}) in the following form
\bea
\nonumber \hat\rho=
\hspace{-0.05cm}\sum_{N_A=0}^N\sum^{N_A}_{\{n_1\dots n_{V_A}\hspace{-0.05cm}\}}\hspace{-0.05cm}
\sum^{N_A}_{\{m_1\dots m_{V_A}\hspace{-0.05cm}\}}
\hspace{-0.cm}\frac{{\cal C}_{N_A}}{\prod_{i=1}^{V_A}n_i! \,m_i!}\\
\nonumber 
\times\,\prod_{i=1}^{V_1}\xi_i^{n_i}b_i^{\dagger n_i}|0_{A_1}\rangle \langle 0_{A_1}|
\prod_{i=1}^{V_1}\xi_i^{*m_i}b_i^{m_i}\\
\otimes
\prod_{j=V_1+1}^{V_A}\xi_j^{n_j}b_j^{\dagger n_j}|0_{A_2}\rangle \langle 0_{A_2}|
\prod_{j=V_1+1}^{V_A}\xi_j^{*m_j}b_j^{m_j}
\eea
so that we can write the partial transpose reduced density matrix $\hat\rho^{T_{A_2}}$ exchanging the partial configurations
$(m_{V_1+1},\dots,m_{V_A})$ and $(n_{V_1+1},\dots,n_{V_A})$. The result matrix is not a block diagonal one anymore. \\
It is much easier considering the homogeneous case in the thermodynamical limit, so that 
\bea
\nonumber
\hat\rho=e^{-V_A\nu}\sum^{\infty}_{n_1\dots n_{V_A}}\hspace{-0.0cm}
\sum^{\infty}_{m_1\dots m_{V_A}}\nu^{\sum_{i=1}^{V_A}{(n_i+m_i)}/{2}}\\
\nonumber \times \,\prod_{i=1}^{V_1}\frac{b_i^{\dagger n_i}}{n_i!}|0_{A_1}\rangle \langle 0_{A_1}|\prod_{i=1}^{V_1}\frac{b_i^{m_i}}{m_i!}\\
\otimes
\hspace{-0.2cm}\prod_{j=V_1+1}^{V_A}\frac{b_j^{\dagger n_j}}{n_j!}|0_{A_2}\rangle \langle 0_{A_2}|\prod_{j=V_1+1}^{V_A}\frac{b_j^{m_j}}{m_j!}
\eea
which is an infinite dimensional sparse matrix. We have, therefore, that, in the homogeneous case, $\xi=\sqrt{\nu/N}$, and for $N,V\rightarrow \infty$,  the partial transpose of the reduced density matrix with respect to a block after a bipartition of the system $A$ becomes simply 
\be
\hat\rho^{T_{A_2}}\rightarrow \hat\rho
\ee
therefore the negativity, which counts the number of negative eigenvalues of $\hat\rho^{T_{A_2}}$, vanishes. This result suggests that, 
in the thermodynamic limit, the state turns to be a so-called entangled PPT (positive partial transpose) state.

\subsection{Mutual Information}
The derivation of the mutual information between the two parts, $A_1$, with size $V_1$, and $A_2$, with size $V_2$, of the system $A=A_1\cup A_2$ after tracing out all the rest, $B$, is very simple. Tracing over $B\cup A_2$ and $B\cup A_1$ we get the entropies $S_{A_1}$ and $S_{A_2}$ respectively.\\ 
We have, therefore, $S$ given by Eq.~(\ref{entropy}) and $S_{A_1}$ and $S_{A_2}$ with the same form as in Eq.~(\ref{entropy}) with $V_1$ and $V_2$, respectively, instead of $V_A$. The mutual information, defined by $I=S_{A_1}+S_{A_2}-S$, is then simply given by
\bea
&&\hspace{-0.1cm} I =\nu\left[V_A \ln V_A-V_1 \ln V_1-V_2 \ln V_2\right]\\
\nonumber&&\hspace{0.2cm}
+\,e^{-V_A\nu}\sum_{n=0}^{\infty}\frac{\nu^n}{n!}\left[e^{V_2\nu}V_1^n+e^{V_1\nu}V_2^n-V_A^n\right]\ln n!
\eea
For large filling fraction $\nu$ or large subsystem sizes $V_1$, $V_2$ (and, therefore, large $V_A=V_1+V_2$) we can use Eq.~(\ref{Sgauss})
for the entropies getting the following simple form for the mutual information
\be
I=\frac{1}{2}\ln\left(\frac{V_1V_2}{V_1+V_2}\right)+\frac{1}{2}\ln(2\pi e\nu)
\ee
Notice that $A$ can be formed by two disjoint parts, $A_1$ and $A_2$, which can be far apart from each other. 
This means that the two subsystems are equally entangled no matter how distant they are. 
\change{We notice that the same behavior occurs for some unconventional quantum spin chains \cite{luca_sc}.}
\section{Correlation functions}
In what follows we show that the diagonal elements of the density matrix $\hat\rho$ are useful to calculate the density correlation functions. We will see that looking only at the density-density correlations, the system of free bosons in the ground state at the thermodynamical limit behaves like a classical uncorrelated gas. On the other hand, the two-site single-particle correlations and, in general, the many-particle correlation functions unveil the off-diagonal long range order and the quantum coherence.
\subsection{Density correlation functions}

Let us first consider the expectation value of $b^\dagger_ib_j$, namely the single-particle correlation function, which, after some algebra can be found to be  
\be
\label{c1}
\langle\Psi| b^\dagger_ib_j|\Psi\rangle=\frac{\xi_i^*\xi_j}{|\xi_j|^2}\,\sum_{n_j=0}^N n_j \rho_{n_j} =N \xi_i^*\xi_j
\ee
where 
\be
\label{binomial}
\rho_{n_j}=\frac{N!}{(N-n_j)!n_j!}(1-|\xi_j|^2)^{N-n_j}|\xi_j|^{2n_j}
\ee
 is the binomial distribution. In particular, defining $\hat n_i=b_i^\dagger b_i$, we have 
 \be
 \label{<n>}
 \langle\Psi| \hat n_i |\Psi\rangle=N |\xi_i|^2
 \ee
 The expectation value of the product of two single particle density operators, $\hat n_i \hat n_j$, namely the density-density correlation function 
 $\langle\Psi| \hat n_i \hat n_j |\Psi\rangle=\langle\Psi| b^\dagger_ib_i b^\dagger_jb_j |\Psi\rangle$, 
 reads
\be
\label{c2}
\langle\Psi| \hat n_i \hat n_j |\Psi\rangle=\hspace{-0.15cm}
\sum_{\{n_i,n_j\}}^{N\,\prime}  n_in_j \,\rho_{n_i n_j}
=(N^2-N) |\xi_i|^2|\xi_j|^2
\ee
for any $i\neq j$ and where 
\be
\label{trinomial}
\rho_{n_i,n_j}=\frac{N!\,(1-|\xi_i|^2-|\xi_j|^2)^{N-n_i-n_j}|\xi_i|^{2n_i}|\xi_j|^{2n_j}}{(N-n_i-n_j)!\,n_i!\,n_j!}
\ee 
is a trinomial distribution. Actually, since we can reduce the order of a multinomial distribution summing over some indices
\be
\sum_{n_i=0}^{N-\sum^\ell_{j\neq i}n_j}\rho_{n_1\dots n_i\dots n_\ell}=\rho_{n_1\dots n_i\hspace{-0.25cm}/\hspace{0.1cm}\dots n_\ell}
\ee
we can use always the reduced density matrix in Eq.~(\ref{rdmD}) to write all the correlation functions choosing arbitrarily a subsystem $A$ which contains the sites involved. For example, for any $i\in [1,{V_A}]$, we can write
\bea
\label{c3}
\nonumber &&
\langle\Psi| \hat n^2_i |\Psi\rangle=
\sum^{N\,\prime}_{\{n_1\dots n_{V_A}\}} 
n^2_i \rho_{n_1\dots n_{V_A}} = \sum^N_{n_i=0} n^2_i \rho_{n_i} \\
&&\phantom{\langle\Psi| \hat n^2_i |\Psi\rangle}
=(N^2-N) |\xi_i|^4+N|\xi_i|^2
\eea
performing the summations of the multinomial distribution, Eq.~(\ref{rdmD}), over all the indices $n_k$ with $k\neq i$ getting, as a result, a binomial  distribution, Eq.~(\ref{binomial}).\\
In general terms, we find that all the density correlation functions, for $k\le {V_A}$, can be written as
\be
\label{correlations}
\langle\Psi| \hat n_1\hat n_2\dots \hat n_k |\Psi\rangle=\sum^{N\,\prime}_{\{n_1\dots n_{V_A}\}} n_1n_2\dots n_k \,\rho_{n_1\dots n_{V_A}}%
\ee
The right-hand-side of Eq.~(\ref{correlations}) is a statistical average, weighted by the multinomial distribution given in Eq.~(\ref{rdmD}), that we will denote by $\langle\dots\rangle_\rho$, so we can write
\be
\label{stat}
\langle\Psi| \hat n_1\hat n_2\dots \hat n_k |\Psi\rangle
=\langle n_1 n_2\dots n_k \rangle_{\rho}
\ee
We find, therefore, that the reduced density matrix in Eq.~(\ref{rdmD}) really plays the role of a statistical distribution so that all the quantum density correlation functions can be seen as related to the moments of such a distribution.\\
In the thermodynamic limit and in the homogeneous case, we can use Eq.~(\ref{poissonD}) so that from Eq.~(\ref{stat}) we can easily calculate all possible density correlation functions exactly, also the expectation values of all possible products of powers of density operators 
\be
\label{Bell}
\langle\Psi| \hat n_1^{\alpha_1} \dots \,\hat n_k^{\alpha_k} |\Psi\rangle=
\langle \hat n_1^{\alpha_1} \dots \,\hat n_k^{\alpha_k} \rangle_\rho=
\prod_{i=1}^kB_{\alpha_i}(\nu)
\ee
where $B_{\alpha_i}(\nu)$ are Bell polynomials, which fulfill the following recurrence relation 
\be
B_{\alpha+1}(\nu)=\sum_{\beta=0}^\alpha \frac{\alpha!}{(\alpha-\beta)! \,\beta!}\,\nu B_{\alpha-\beta}(\nu)
\ee
with initial value $B_0(\nu)=1$. 
For example, $B_1(\nu)=\nu$, $B_2(\nu)=(\nu^2+\nu)$, $B_3(\nu)=(\nu^3+3\nu^2+\nu)$, etc.\\
Before we proceed a comment is in order. We saw that, for the homogeneous case and in the thermodynamic limit, all the density correlation functions factorize, Eq.~(\ref{Bell}). This means, in particular, that the covariance matrix is diagonal, namely the two-point density-density connected correlation functions vanish when we go to the thermodynamic limit, 
$\forall \,i\neq j$, 
\be
\lim_{N,V\rightarrow \infty}\big(\langle \Psi|n_i n_j|\Psi\rangle-\langle \Psi|n_i|\Psi\rangle \langle \Psi|n_j|\Psi\rangle\big)= 0
\ee
which means that the particles become uncorrelated. 

\bigskip
\paragraph{Particle number fluctuations.}
Using Eqs.~(\ref{<n>}), (\ref{c2}) and (\ref{c3}), we can calculate the fluctuations of the total number of particles in the subsystem $A$. Calling $\hat N_A=\sum_{i=1}^{{V_A}}\hat n_i$ we have that the variance of this number is 
\bea
\label{nf}
\nonumber&&\hspace{-0.35cm}
\delta N_A=\langle\Psi|\hat N_A^2 |\Psi\rangle-\langle\Psi|\hat N_A|\Psi\rangle^2=\\
\nonumber &&\hspace{-0.3cm}\sum_{i=1}^{{V_A}}\Big(\langle\Psi|\hat n^2_i|\Psi\rangle+2\sum_{j\neq i}^{V_A}\langle\Psi|\hat n_i\hat n_j|\Psi\rangle\Big)-\Big(\sum_{i=1}^{{V_A}}\langle\Psi|\hat n_i|\Psi\rangle\Big)^2\\
&&\hspace{-0.25cm} =N\Big[\sum_{i=1}^{V_A} |\xi_i|^2-\Big(\sum_{i=1}^{V_A} |\xi_i|^2\Big)^2\Big]
\eea
For an homogeneous system on a torus we have $|\xi_i|^2=\frac{1}{V}$ and Eq.~(\ref{nf}) becomes simply
\be
\delta N_A={V_A} \,\nu\left(1-\frac{{V_A}}{V}\right)
\ee
where we used Eq.~(\ref{filling}). In the thermodynamic limit, namely for $V\rightarrow \infty$, therefore, the fluctuations increase linearly with the volume ${V_A}$ of the subsystem $A$, consistently with the behavior of the entropy in Eq.~(\ref{entropy}) in the very dilute regime, $\nu\ll 1$. 
\changes{We clarified by this analysis how to reconcile the logarithmic behavior of the entanglement entropy with the volume law of the particle number fluctuations by means of a unified approach through the reduced density matrix.}

\subsection{Generating function}
In order to calculate all the density correlation functions it is convenient to introduce the so called moment generating function ${\cal G}$, introducing  auxiliary source fields $\mu_i$ such that any density correlators can be obtained by deriving many times ${\cal G}$ with respect to 
$\mu_i$. This quantity can be calculated as it follows
\be
{\cal G}(\{\mu_i\})=\sum^{N\,\prime}_{\{n_1\dots n_{V_A}\}} 
\rho_{n_1\dots n_{V_A}} e^{\sum_i\mu_i n_i} 
\ee 
which, explicitly, reads  
\bea
\label{G}
\nonumber {\cal G}(\{\mu_i\})=\sum^{N\,\prime}_{\{n_1\dots n_{V_A}\}} 
 \frac{N!\,\Big(1-\sum_{i=1}^{V_A} |\xi_i|^2\Big)^{N-N_A}
}{\left(N-N_A\right)!}\\
\times\,\prod_{i=1}^{V_A}\frac{|\xi_i|^{2n_i}e^{\mu_i n_i} }{n_i!}\hspace{1cm}
\eea
Performing the summation we easily get
\be
{\cal G}(\{\mu_i\})=\left[\Big(1-\sum_{i=1}^{V_A} |\xi_i|^2\Big)+\sum_{i=1}^{V_A}|\xi_i|^{2}e^{\mu_i}\right]^N
\ee
Making the derivatives of this function with respect to $\mu_i$ one can easily get all the density correlation functions. For instance, we have
\be
\label{Gderiv}
\langle\Psi| n_1^{\alpha_1}\dots \, n_k^{\alpha_k}|\Psi\rangle=
\left.\frac{\partial^{\sum_{i}^k\alpha_i} \,{\cal G}}{\partial \mu_1^{\alpha_1} \dots \partial \mu_k^{\alpha_k} }\right|_{\mu_i=0}
\ee
which is the generalization of Eq.~(\ref{Bell}) to the inhomogeneous system. For the homogeneous case Eq.~(\ref{G}) reduces to
\be
{\cal G}(\{\mu_i\})=\Big(1-\frac{V_A\nu}{N}+\frac{\nu}{N}\sum_{i=1}^{V_A}e^{\mu_i}\Big)^N
\ee
which, in the thermodynamic limit, namely for $N\rightarrow \infty$, becomes simply
\be
\label{Gh}
{\cal G}(\{\mu_i\})=\exp\Big[{\nu\Big(\sum_{i=1}^{V_A}e^{\mu_i}-V_A\Big)}\Big]
\ee
Taking the derivatives with respect to $\mu_i$ and the limit $\mu_i\rightarrow 0$ we generate the Bell polynomials, so that using Eq.~(\ref{Gh}) in Eq.~(\ref{Gderiv}) we obtain Eq.~(\ref{Bell}).

\subsection{Pair correlation functions}
We have seen already that the one-particle correlation functions, Eq.~(\ref{c1}), are finite. Let us now consider the so-called pair correlation functions which describe the process for a couple of particles to make a quantum hopping from one site to another. From 
\bea
\nonumber 
b_jb_j|\Psi\rangle=\hspace{-0.1cm}\sum^N_{\{n_1,\dots n_V\}}\hspace{-0.1cm} n_j(n_j-1)
\frac{\sqrt{N!}}{\prod_{i=1}^V n_i!}\xi_j^{n_j}b_j^{\dagger n_j-2}\\
\times \prod_{k\neq j}^{V}\xi_k^{n_k}
b_k^{\dagger n_k}  |0\rangle
\label{b2}
\eea
after some algebraic steps we get the following expression for the pair correlation functions
\be
\langle\Psi| b_i^\dagger  b_i^\dagger b_jb_j  |\Psi\rangle=\frac{\xi_i^{*2}\xi_j^2}{|\xi_j|^4}
\langle n_j(n_j-1)\rangle_\rho
\ee
and using Eqs.~(\ref{<n>}) and (\ref{c3}) we find
\be
\label{pair}
\langle\Psi| b_i^\dagger  b_i^\dagger b_jb_j  |\Psi\rangle={\xi_i^{*2}\xi_j^2}(N^2-N)
\ee
Eq.~(\ref{c1}) and Eq.~(\ref{pair}), valid for any couple of points $(i,j)$ are the manifestation of the off-diagonal long-range order in the system. For the homogeneous case we have $|\xi_i|^2=\nu/N$ so that, in the thermodynamic limit, Eq.~(\ref{pair}) reduces to
\be
\label{b2b2}
\langle\Psi| b_i^\dagger  b_i^\dagger b_jb_j  |\Psi\rangle=\nu^2
\ee
This result can be generalized as shown in the next section. 
\changes{For completeness we can calculate analogously the four-point correlation functions, which turn out to be
\be
\langle\Psi| b_i^\dagger  b_k^\dagger b_jb_\ell  |\Psi\rangle=\frac{\xi_i^{*}\xi_k^{*}\xi_j\xi_\ell}{|\xi_j|^2|\xi_\ell|^2}\langle n_j n_\ell\rangle_\rho.
\ee
Using Eq.~(\ref{c2}), we can write
\be
\label{4p}
\langle\Psi| b_i^\dagger  b_k^\dagger b_jb_\ell  |\Psi\rangle={\xi_i^{*}\xi_k^{*}\xi_j\xi_\ell}(N^2-N)
\ee
which, in the homogeneous case and in the thermodynamic limit, is equal to Eq.~(\ref{b2b2}). 
Finally we notice that these correlation functions do not factorize according to the contraction prescriptions of Wick's theorem.
}

\subsection{Many-particle correlation functions}
We generalize the single-particle and the pair correlation functions considering the many-particle correlation functions. For any integer $\alpha$ we find that these two-point correlation functions are given by the following equation
\be
\langle\Psi| b_i^{\dagger\alpha} b_j^\alpha |\Psi\rangle=\frac{\xi_i^{*\alpha}\xi_j^\alpha}{|\xi_j|^{2\alpha}}\Big\langle\frac{n_j!}{(n_j-\alpha)!}\Big\rangle_\rho
\label{pair^a<>}
\ee
which, performing the summation, reads
\bea
\nonumber 
\langle\Psi| b_i^{\dagger\alpha} b_j^\alpha |\Psi\rangle=\frac{_2F_1\big(-N,1,1-\alpha;\frac{|\xi_j|^2}{|\xi_j|^2-1}\big)}{\Gamma(1-\alpha)}\\
\times\, (1-|\xi_j|^2)^N\,\frac{\xi_i^{*\alpha}\xi_j^\alpha}{|\xi_j|^{2\alpha}}
\label{pair^a}
\eea
where $\frac{_2F_1(a,b,c;z)}{\Gamma(c)}$ is a regularized hypergeometric function. For the homogeneous case and in the thermodynamic limit Eq.~(\ref{pair^a}) simplifies as follows
\be
\label{baba}
\langle\Psi| b_i^{\dagger\alpha} b_j^\alpha |\Psi\rangle=\nu^\alpha
\ee
This quantity describes the quantum hopping of $\alpha$ particles from the site $j$ to the site $i$. The composite operator ${\cal B}^{(\alpha)}= b_i^{\dagger\alpha} b_j^\alpha$, therefore, acts within the same sector of fixed number of particles, exchanging two configurations of the occupation numbers, 
\[(n_1,\dots n_i,\dots n_j,\dots) \rightarrow (n_1,\dots (n_i+\alpha),\dots (n_j-\alpha),\dots)\]
In other words, any correlation function of the type $\langle\Psi| b_i^{\dagger\alpha} b_j^\alpha |\Psi\rangle$ can be written as the trace of the product of a reduced density matrix $\hat \rho$ times an operator ${\cal B}^{(\alpha)}$, off-diagonal in the configuration space, with elements
\be
\label{B}
{\cal B}^{(\alpha)}_{{\substack{m_1\dots m_{V_A}\\n_1\dots n_{V_A}}}}
\hspace{-0.1cm}=\sqrt{\frac{(n_i+\alpha)!n_j!}{n_i!(n_j-\alpha)!}}\delta_{m_i,n_i+\alpha}\delta_{m_j,n_j-\alpha}
\hspace{-0.1cm}\prod_{k\neq i,j} \delta_{m_k,n_k}
\ee
The two-point many-particle correlation functions can be written, therefore, as follows
\be
\label{TrBrho}
\langle\Psi| b_i^{\dagger\alpha} b_j^\alpha |\Psi\rangle=\Tr(\hat \rho\,{\cal B}^{(\alpha)})
\ee
which is a finite quantity only because of the off-diagonal coherence terms in $\hat\rho$. Actually, putting Eqs.~(\ref{rdm}) and (\ref{B}) in Eq.~(\ref{TrBrho}) 
\[
\Tr(\hat \rho\,{\cal B}^{(\alpha)})=\hspace{-0.05cm}\sum_{N_A=0}^N\sum^{N_A}_{\{n_1\dots n_{V_A}\hspace{-0.05cm}\}}\hspace{-0.05cm}
\sum^{N_A}_{\{m_1\dots m_{V_A}\hspace{-0.05cm}\}}
\rho_{\substack{n_1\dots n_{V_A}\\m_1\dots m_{V_A}}} 
{\cal B}^{(\alpha)}_{{\substack{m_1\dots m_{V_A}\\n_1\dots n_{V_A}}}}
\]
we obtain Eq.~(\ref{pair^a<>}) and, therefore, Eq.~(\ref{pair^a}). Notice that the correlation functions written in Eq.~(\ref{TrBrho}) manifestly depend on the off-diagonal elements of the density matrix $\hat\rho$, although in Eq.~(\ref{pair^a<>}) they are written in terms of the diagonal elements. 

\subsection{Phase fluctuations}
As a final remark, let us introduce a space-dependent phase,   
\changes{possibly caused by the presence of an inhomogeneity of the system,}  in the single particle wavefunction  
parametrized by a space-dependent modulus and angle
\be
\xi_j=z_j e^{i\theta_j}
\ee
where $z_j$ is a real number, so that, contrary to the density correlation functions which always depend on $|\xi_i|^2$, the many-particle correlation functions in Eq.~(\ref{pair^a}) will depend on the phase
\be
\label{phase_many}
\langle\Psi| b_i^{\dagger\alpha} b_j^\alpha |\Psi\rangle\propto e^{i\alpha(\theta_j-\theta_i)}
\ee
At the same time, also the off-diagonal elements of the density matrix in Eq.~(\ref{rdm}) are phase dependent, providing that the phase is not uniform, in the following way
\be
\label{phase_rdm}
\rho_{\substack{n_1\dots n_{V_A}\\m_1\dots m_{V_A}}} \propto e^{i\sum_{i=1}^{V_A} (n_i-m_i)\theta_i} 
\ee
with the constraint $\sum_{i=1}^{V_A} n_i=\sum_{i=1}^{V_A} m_i$, therefore for constant $\theta_i=\theta$  the phase dependence cancels out. \\
For large phase fluctuations, the many-particle hopping described by Eq.~(\ref{phase_many}) can be suppressed within the stationary phase approximation, as well as the off-diagonal coherence terms of the reduced density matrix, Eq.~(\ref{phase_rdm}).

This observation \changes{suggests} that, after a bipartition, if one could introduce a wide space modulation of the phase or a random phase field, for instance, by breaking time reversal symmetry so that $t_{ij}\rightarrow t_{ij}e^{i(\theta_i-\theta_j)}$ with $\theta_i$ random variables, one could spoil the coherence and the quantum fluctuations, getting, at the same time, a diagonal reduced density matrix. As a result, we would get an extensive entropy fulfilling a volume law, but with the drawback of a vanishing mutual information. The Von Neumann entropy, therefore, would resemble a ``thermal" entropy of a gas of classical particles in a grand canonical ensemble.

\section{Conclusions}
We calculate analytically the reduced density matrix in the configuration space, the entanglement entropy, the R\`enyi entropies, the quantum negativity and the mutual information between two separated regions for an extended system of free bosons.
We show that, after a bipartition, the mixed state described by the reduced density matrix in the thermodynamic limit is an entangled PPT state, with vanishing negativity but finite space-independent mutual information. 

Moreover we show that the reduced density matrix written in terms of occupation numbers can be seen as a statistical distribution useful to derive all the particle correlation functions. 
We find that in the thermodynamic limit, looking only at the density-density correlation functions, the system of free bosons behaves like an uncorrelated gas since all the density operators are diagonal in the configuration space and can be calculated exactly at any order from the moment generating function of a multinomial distribution. 
The coherence among the particles, instead, generates off-diagonal terms in the reduced density matrix and off-diagonal long-range order described by always finite many-particle correlation functions. Finally we discussed how, within this description, a random field might spoil the quantum fluctuations.

\subsection{Acknowledgements}
The author acknowledges financial support from the project BIRD 2021 ``Correlations, dynamics and topology in long-range quantum systems" of the Department of Physics and Astronomy, University of Padova.


%
\section{Appendix}

Here we present some details of the calculations reported in the main text. 
\change{
\paragraph{Ground state.}
Assuming $t_{ij}$ a hermitian matrix which can be diagonalized by a unitary transformation $U$, 
\be
-\sum_{ij} U_{ki}t_{ij}U^\dagger_{jq}=\epsilon_k\delta_{kq}
\ee
we can define the canonically transformed fields
\be
\tilde b_{k}=\sum_{i}U_{ki} b_i
\ee
since $U$ preserves the commutation relations. The Hamiltonian in Eq. (\ref{H1}) becomes 
\be
H=\sum_{k}\epsilon_k\, \tilde b^\dagger_k \tilde b_k.
\ee
At zero temperature the lowest energy state denoted by $k=k_0$ hosts all the $N$ particles, forming a condensate, so that, calling 
$\xi_i\equiv U^*_{k_0 i}$, the ground state is given by
\be
|\Psi\rangle=\frac{1}{\sqrt{N!}}\Big(\tilde b^\dagger_{k_0}\Big)^N|0\rangle=\frac{1}{\sqrt{N!}}\Big(\sum_{i}\xi_i b^\dagger_i\Big)^N|0\rangle
\ee
reported in Eq. (\ref{state}). 
For a translationally invariant system $U_{k j}=\frac{1}{\sqrt{V}}e^{i k\cdot j}$, namely it is the Fourier transformation which diagonalizes the Hamiltonian. The lowest energy state is at $k=0$, so that we have $\xi_i=U^*_{0i}=1/\sqrt{V}$.
}
\paragraph{Reduced density matrix.} 
For the reduced density matrix, Eq.~(\ref{rdmA}), we need to evaluate
\be
\langle 0_B| \prod_{i={V_A}+1}^{V}b_i^{m_i}|\Psi\rangle
\ee
where $|\Psi\rangle$ is given in Eq.~(\ref{state}), therefore, after splitting the two subsystems $A$ and $B$, it can be written as
\bea
\nonumber
\langle 0_B| \prod_{i={V_A}+1}^{V}b_i^{n_i}|\Psi\rangle=\hspace{-0.15cm}
\sum^N_{\{n_1\dots n_V\}}\frac{\sqrt{N!}}{\prod_{i=1}^V n_i!}
\prod_{j=1}^{V_A}\xi_j^{n_j}b_j^{\dagger{n_j}}|0_A\rangle\\
\nonumber
\times\, \langle 0_B| \prod_{i={V_A}+1}^{V}b_i^{m_i}\hspace{-0.1cm}
\prod_{j=V_A+1}^{V}\xi_j^{n_j}b_j^{\dagger{n_j}}|0_B\rangle  \hspace{0.2cm}\\
=\sqrt{N!}\prod_{j=V_A+1}^{V}\xi_j^{n_j}\hspace{-0.1cm}
\sum^N_{\{n_1\dots n_{V_A}\}}
\prod_{i=1}^{V_A}\frac{\xi_i^{n_i}}{n_i!}b_i^{\dagger{n_i}}|0_A\rangle \hspace{0.2cm}
\eea
where we used
\be
\label{bb}
\langle 0| b_i^{m} b_j^{\dagger n}|0\rangle=\delta_{n m}\,\delta_{i j}\,n!
\ee
For the trace over $B$ we used the following relation (let us call $V_A'=V_A+1$ for brevity)
\be
\sum^{N-N_A}_{\{n_{V_A'}\dots n_V\}} \prod_{j=V_A'}^{V}\frac{|\xi_j|^{2n_j}}{n_j!}=
\frac{(\sum_{j=V_A'}^V|\xi_j|^2)^{N-N_A}}
{(N-N_A)!}
\ee
getting the final result in Eq.~(\ref{rdmA2}). 

\paragraph{Correlation functions.}
From Eq.~(\ref{state}), applying an annihilation operator we get
\be
b_j|\Psi\rangle=\hspace{-0.2cm}\sum^N_{\{n_1,\dots n_V\}}\hspace{-0.2cm} n_j
\frac{\sqrt{N!}}{\prod_{i=1}^V n_i!}\xi_j^{n_j}b_j^{\dagger n_j-1}
\prod_{k\neq j}^{V}\xi_k^{n_k}
b_k^{\dagger n_k}  |0\rangle
\ee
and using again the identity Eq.~(\ref{bb}), after some algebraic steps we get
\bea
\nonumber &&\hspace{-0cm}\langle\Psi| b^\dagger_ib_j|\Psi\rangle=N!\hspace{-0.2cm}
\sum^{N}_{\{n_1\dots n_V\}}\prod_{k=1}^V\frac{|\xi_k|^{2n_k}}{n_k!}\,n_j\frac{\xi_i^\dagger}{\xi_j^\dagger}\\
\nonumber&&=N!\hspace{-0.05cm} \sum_{n_j=0}^N\sum^{N-n_j}_{\{n_1..n_j{\hspace{-0.26cm}/}\;.. n_V\}}
\prod_{k\neq j}^V\frac{|\xi_k|^{2n_k}}{n_k!}
\frac{|\xi_j|^{2n_j}}{n_j!}n_j\frac{\xi_i^\dagger}{\xi_j^\dagger}\\
 \nonumber &&=N!\hspace{-0.05cm} \sum_{n_j=0}^N\frac{(\sum_{k\neq j}|\xi_k|^2)^{N-n_j}|\xi_j|^{2n_j}}{(N-n_j)!\,n_j!}n_j\,
\frac{\xi_i^\dagger \xi_j}{|\xi_j|^2}\\
&&=\sum_{n_j=0}^N n_j \rho_{n_j}\, \frac{\xi_i^*\xi_j}{|\xi_j|^2}=N \xi_i^*\xi_j
\eea
where $\rho_{n_j}$ is the binomial distribution defined in Eq.~(\ref{binomial}). 
Let us now calculate the density-density correlation functions making the scalar product of two vectors of the form
\be
b_j^\dagger b_j|\Psi\rangle=\hspace{-0.1cm}\sum^N_{\{n_1,\dots n_V\}}\hspace{-0.1cm} n_j
\frac{\sqrt{N!}}{\prod_{i=1}^V n_i!}\prod_{k\neq j}^{V}\xi_k^{n_k}\xi_j^{n_j}
b_k^{\dagger n_k} b_j^{\dagger n_j} |0\rangle
\ee
so that, after analogous algebraic steps, we have
\bea
\nonumber &&\langle\Psi|\hat n_i\hat n_j|\Psi\rangle=N!\hspace{-0.2cm}\sum^{N}_{\{n_1\dots n_V\}}\prod_{k=1}^V\frac{|\xi_k|^{2n_k}}{n_k!}\,n_in_j\\
\nonumber &&=N!\hspace{-0.05cm} \sum_{\{n_i,n_j\}}^{N\,\prime}\frac{(\sum_{k\neq i,j}|\xi_k|^2)^{N-n_i-n_j}|\xi_i|^{2n_i}|\xi_j|^{2n_j}}{(N-n_i-n_j)!\,n_i!n_j!}n_in_j\\
&&=\sum_{\{n_i,n_j\}}^{N\,\prime}  n_in_j \,\rho_{n_i n_j}
\eea
where $\rho_{n_i n_j}$ is the trinomial distribution  in Eq.~(\ref{trinomial}) and the sum means $\sum_{\{n_i,n_j\}}^{N\,\prime}=\sum_{n_i=0}^{N}\sum_{n_j=0}^{N-n_i}$.\\ One can easily verify that, in general 
\bea
\nonumber\langle\Psi| \hat n_1\hat n_2\dots \hat n_k |\Psi\rangle \hspace{-0.1cm}&=&\hspace{-0.1cm} 
N!\hspace{-0.2cm}\sum^{N}_{\{n_1\dots n_V\}}\prod_{i=1}^V\frac{|\xi_i|^{2n_i}}{n_i!}\,n_1n_2\dots n_k\\
\hspace{-0.1cm}&=&\hspace{-0.1cm}\sum^{N\,\prime}_{\{n_1\dots n_{k}\}} n_1n_2\dots n_k \,\rho_{n_1\dots n_{k}}
\eea
where $\rho_{n_1\dots n_{k}}$ is a multinomial distribution of order $k$. 
We can show now that we can reduce the order of a multinomial distribution summing over some indices, for instance
\bea
\nonumber && \hspace{-0.5cm}
\sum_{n_i=0}^{N-\sum^\ell_{j\neq i}n_j}\rho_{n_1\dots n_i\dots n_\ell}=N!\prod_{j\neq i}^\ell\frac{|\xi_j|^{2n_j}}{n_j!} \hspace{-0.cm}\\
\nonumber &&\times \hspace{-0.2cm}\sum_{n_i=0}^{N-\sum^\ell_{j\neq i}n_j}\hspace{-0.1cm}
\frac{\big(1-\sum_{j=1}^\ell |\xi_j|^2\big)^{N-\sum_{j=1}^\ell n_j}|\xi_j|^{2n_i}}{(N-\sum_{j=1}^\ell n_j)!\,n_i!}\\
\nonumber &&=\frac{N! \big(1-\sum_{j\neq i}^\ell |\xi_j|^2\big)^{N-\sum_{j\neq i}^\ell n_j}\prod^\ell_{j\neq i}|\xi_j|^{2n_j}}{(N-\sum^\ell_{j\neq i}n_j)! \prod_{j\neq i}^\ell n_j! }\\
&&=\rho_{n_1\dots n_i\hspace{-0.25cm}/\hspace{0.1cm}\dots n_\ell}
\eea
where in the second equation we recognize the sum of a binomial distribution with $(N-\sum_{j\neq i}^\ell n_j)$ total number of particles.
In this way we can use just a single multinomial distribution, Eq.~(\ref{rdmD}), to write all the density correlation functions in terms of its moments, providing that the subset $A$ contains all the sites involved, as reported in Eq.~(\ref{correlations}) and Eq.~(\ref{stat}), where we defined
\be
\langle n_1 n_2\dots n_k \rangle_{\rho}=\sum^{N\,\prime}_{\{n_1\dots n_{V_A}\}} n_1n_2\dots n_k \,\rho_{n_1\dots n_{V_A}}
\ee

\noindent For example the variance and the covariance (for $i\neq j$) of the multinomial distribution Eq.~(\ref{rdmD}) are 
\bea
\textrm{Var}(n_i)=
\langle n^2_i\rangle_{\rho}-\langle n_i\rangle^2_{\rho}=N|\xi_i|^2(1-|\xi_i|^2)\;\;\;\;\;\,\\
\textrm{Cov}(n_i,n_j)= 
\langle n_in_j\rangle_{\rho}-\langle n_i\rangle_{\rho}\langle n_j\rangle_{\rho} =-N|\xi_i|^2|\xi_j|^2
\eea
For the homogeneous case $|\xi_i|^2=1/V=\nu/N$ and in the thermodynamic limit, $V,N\rightarrow \infty$, but finite $\nu=N/V$, 
Eq.~(\ref{rdmD}) simplifies to Eq.~(\ref{poissonD}) so that, for instance, $\textrm{Var}(n_i)=\nu$, $\textrm{Cov}(n_i,n_j)=0$ and in general terms, since
\be
\sum_{n=0}^\infty \frac{\nu^n}{n!} \,n^\alpha=e^{\nu} B_\alpha(\nu) 
\ee
where $B_\alpha(\nu) $ are Bell polynomials, we can easily verify the result for any kind of density correlation functions reported in Eq.~(\ref{Bell}).

For the many-particle correlation functions we have to apply many times $b_j$ on the ground state 
\bea
\nonumber 
b_j^\alpha|\Psi\rangle=\hspace{-0.1cm}\sum^N_{\{n_1,\dots n_V\}}\hspace{-0.1cm} \frac{n_j!}{(n_j-\alpha)!}
\frac{\sqrt{N!}}{\prod_{i=1}^V n_i!}\xi_j^{n_j}b_j^{\dagger n_j-\alpha}\\
\times \prod_{k\neq j}^{V}\xi_k^{n_k}
b_k^{\dagger n_k}  |0\rangle
\eea
so that, making the scalar product with another vector of the same kind, and using Eq.~(\ref{bb}), we get
\be
\langle\Psi| b_i^{\dagger\alpha} b_j^\alpha |\Psi\rangle=\frac{\xi_i^{*\alpha}}{\xi_j^{*\alpha}}
N!\hspace{-0.2cm}\sum^{N}_{\{n_1\dots n_V\}}\frac{n_j!}{(n_j-\alpha)!}
\prod_{k=1}^V\frac{|\xi_k|^{2n_k}}{n_k!}
\ee
which can be recast in terms of the multinomial distribution as in Eq.~(\ref{pair^a<>}).
%
 

\end{document}